# Approaches to the Algorithmic Allocation of Public Resources: A Cross-disciplinary Review

Saba Esnaashari, Jonathan Bright, John Francis, Youmna Hashem, Vincent Straub, Deborah Morgan

**Abstract**

Allocation of scarce resources is a recurring challenge for the public sector: something that emerges in areas as diverse as healthcare, disaster recovery, policing, and social welfare. The complexity of these policy domains and the need for meeting multiple and sometimes conflicting criteria has led to increased focus on the use of algorithms in this type of allocation decision, which has prompted a wide variety of research across a diverse array of sectors. However, little engagement between researchers across these domains has happened, meaning a lack of understanding of common problems and techniques for approaching them. Here, we performed a cross disciplinary literature review to understand approaches taken for different areas of algorithmic allocation including healthcare, organ transplantation, homelessness, disaster relief, and welfare. We initially identified 1070 papers by searching the literature, then six researchers went through them in two phases of screening resulting in 176 and 75 relevant papers respectively. We then analyzed the 75 papers from the lenses of optimization goals, techniques, interpretability, flexibility, bias, ethical considerations, and performance. We categorized approaches into human-oriented versus resource-oriented perspectives, and individual versus aggregate. We identified that 76% of the papers approached the problem from a human perspective and 60% from the aggregate level using optimization techniques. We found considerable potential for performance gains, with optimization techniques often decreasing waiting times and increasing success rate by as much as 50%. However, we also found a lack of attention to responsible innovation: only around one third of the papers considered ethical issues in choosing the optimization goals while just a very few of them paid attention to the bias issues. Our work can serve as a guide for policy makers and researchers wanting to use an algorithm for addressing a resource allocation problem, and also points to potential future research directions in the field.

## Introduction

Allocation of scarce resources is the primary focus of economics where the objective is to allocate resources in a manner to reach optimal social welfare or an (Nash) equilibrium. However, in the case for free public resources or social provision for vulnerable groups such as state supported housing, child welfare, or healthcare, the problem of achieving an optimum allocation is complex and challenging (Freedman et al. 2020, Kube et. al. 2022, Das 2022).

With the advance of AI and the introduction of a vast set of algorithms, new avenues have arisen for automating the process of prioritizing people for receiving resources. Automation can potentially save time and resources; using algorithms for the problem also has the potential of mitigating human bias and increasing fairness in allocation (Mullainathan 2019); though of course algorithms are not 'fair' by default and bias can easily creep in during their construction and deployment. Furthermore, the criteria needed for making prioritization decisions, the features for finding the most entitled individuals, alongside fairness and efficiency definitions depends on the context, time, and the society's perception from the notion of local justice as named by Elster (1992) and stated by Das (2022). In this context, automated systems can be leveraged to engage human participation in the process of design, and that can facilitate democracy along other benefits of algorithms. In the context of organ transplantation for example, Papalexopoulos et. al. (2022) developed a flexible tool using machine learning and optimization techniques to predict outcomes upon multiple and conflicting objectives allowing policy makers to make the trade-offs between fairness and efficiency based on the society's preferences.

Considering the benefits of using algorithms for allocating social resources, we performed a literature review to understand the strength and weakness of the research in different areas where individuals or households should be prioritized for receiving resources. Our areas of study include healthcare (triage, scheduling, pandemic response), disasters, organ transplantation, homelessness, and welfare. We analyzed 75 papers from the lenses of optimization goals, techniques, performance, interpretability, flexibility, bias, and ethical considerations. We identified that 76% of the papers have considered problem from a human-oriented perspective while others from a resource-oriented or a combination of human and resource. In addition, 60% of the papers approached the problem from aggregate level formulating it as an optimization model while the rest considering individual level using machine learning techniques. Detailed explanations will come in the analysis section.

Alongside the benefits of this paper for scholars, we hope our work helps policy makers understand the research currently done in this scope and view some potential avenues for bringing the findings of these papers into practice. In the next sections, we briefly describe the methodology for reviewing the papers. Then, we will analyze the literature from the lens of each metric as shown in Table 1. We conclude with a discussion of our results and their implications for the field at large.

## Methodology

Taking insight from the Preferred Reporting Items for Systematic Reviews and Meta-Analysis (PRISMA) protocol (Moher et. al. 2015), we went through the literature for algorithmic allocation of public resources to the individuals in three phases: Identification, Screening and Metric selection.

### Identification

Our main objective was to select papers which used some kind of algorithm for making allocation decisions, or prioritizing individuals for receiving scarce resources. Hence, we searched Scopus with the combination of the following keywords: ("resource allocation" or "prioritization") and ("algorithm" or "Artificial Intelligence" or "machine learning" or "decision support system") in the scope of ("patient scheduling" or "triage" or "medical care" or "disaster" or "organ transplantation" or "food" or "homelessness" or "welfare") in the period of 2020-2023. This resulted in 1,070 papers for the screening phase.

### Screening

In the screening phase, six members of our research team went through 1,070 papers in three different rounds. During the first round, 15 papers were selected randomly from the 1,070 and a reliability check was carried out to tune our perception of the papers' relevance based on the inclusion and exclusion criteria, and their type, as defined in Text box 1. In the second round, the 1,070 papers were randomly divided and assigned to the team, with each member responsible for screening between 130-200 papers based on the title and the abstract. This exercise led to the inclusion of 30 review papers, 43 guideline papers, and 176 application papers. In line with the aims of this review, our analysis focuses specifically on application papers. In the final round of the screening, the 176 application papers were reassigned between the six members for a final reliability/validation check, with each member assigned between 18-35 papers to screen based on the whole text of the paper. Following this third and final round of screening, a total of 75 relevant application papers were identified for analysis.

### Metric Selection

The metric selection phase was an iterative, multi-step process that occurred in tandem with previous phases. Following the existent metrics widely used in the field of computer science we identified a set of metrics related to the technical dimensions. Then, to capture the socio-technical elements shaping these systems, we referred to key literature exploring the dimensions of resource allocation in the context of local justice (Das 2022). An additional two members with a disciplinary background in ethics and responsible design of AI systems were also brought onto the research team to guide the development of the metrics. Finally, taking insight from Straub et. al. (2022), we categorised metrics into four different domains of thematic, operational (design and performance), epistemic and normative and identified some metrics for evaluation in each category which is summarised in Table 1.

Next, we will analyse the literature from the lens of each metric. In addition, further rationale is provided for the inclusion of each metric in the analysis section.

| Domain | Metric | Definition |
|---|---|---|
| Thematic | Scope | The main scope and sub-scope for the resource allocation problem. |
| Operational (Design) | Perspective | The overall aim of the allocation |
| Operational (Design) | Approach and Technique | Aggregate versus Individual approach and the algorithm used for allocation decisions. |
| Operational (Design) | Optimization Goal or prioritization metric | The prioritization metric or the main objective of the optimization model. |
| Epistemic | Interpretability | Mentioning interpretability or any concern around interpretability |
| Epistemic | Flexibility | If the design of the system is flexible based on multiple different metrics. |
| Normative | Ethical Considerations | Whether authors have brough any ethical reasoning for considering the prioritisation metric |
| Normative | Bias | Whether bias is detected or mitigated |
| Operational (performance) | Evaluation | Approach to evaluation chosen including Accuracy, Precision, Recall, AUC, F1 Score, Entropy, etc and the value achieved |
| Operational (performance) | Comparison to Status Quo | Waiting time, utilization, survival rate, etc |
| Operational (performance) | | Traditional System values / New System values |

Table 1: Metrics chosen for analysis.

> **Inclusion Criteria**
> - Using some kind of algorithm for the purpose of allocating resources which is provided to the individuals.
>
> **Exclusion Criteria**
> - Using algorithm for allocation of tasks for computing e.g. scheduling packets for transfer across the internet
> - Allocation of resources to the departments rather than individuals e.g., allocation of beds to the hospitals.
> - Using some kind of algorithm for prediction about individuals which helps for a better prioritization but not doing a direct allocation.
>
> **Type of Relevant Papers**
> **Review**
> A literature review for algorithmic resource allocation in general or any specific scope as described in the text.
> **Guideline**
> Unifying guidelines and criteria for allocation of resources (using algorithms) through investigating the literature or doing a survey from the experts.
> **Application**
> Implementing an algorithm for the sake of allocating resources to the individuals in the areas shown in figure 1.

Text box 1: Inclusion/Exclusion Criteria and Review/guideline/application definitions.

## Analysis

In this section, we will go over each metric as listed in Table 1 and analyze the papers from the lens of that metric.

### Scope

We identified healthcare, disaster, organ transplantation, homelessness, and welfare as the domains where individuals or households should be prioritized for receiving resources. Healthcare consisted of the most focused area with 71% of the papers followed by disaster and organ transplantation with 14% and 9% of the papers respectively. Homelessness and welfare were the two least noticed scopes. Perhaps this seems reasonable according to the greater number of people in the healthcare, disaster and organ transplantation compared to the homelessness and welfare (figure 1).

Within healthcare, patient scheduling or operating room scheduling was the most focused category followed by triage and covid. Within organ transplantation kidney was the most widely focused category. Within disaster, most papers didn't mention any specific type. However, between those that had mentioned a particular type, earthquakes and maritime consisted of the two most popular areas. For more details see table A1 in the appendix.

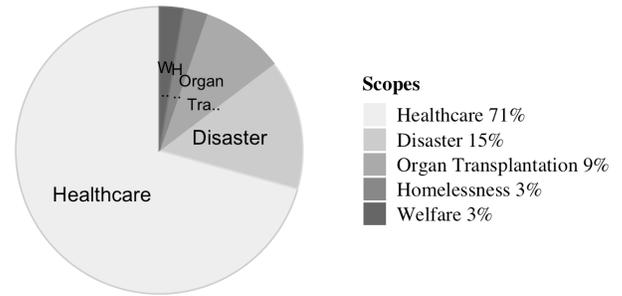

Figure 1: Scopes of Algorithmic Resource Allocation

### Perspective: Human-Oriented vs. Resource-Oriented

We divide up the aim of the allocation algorithm into two types: human oriented and resource oriented. For example, the problem of bed allocation in the hospitals during a pandemic can be modelled in two different ways. In the first setting the problem can be modelled as, given a certain number of beds, what would be the optimal allocation to minimize waiting time or increase survival rate of the patients? In the second setting, the problem can be modelled as given a certain threshold for the waiting time or survival rate, what would be the optimal allocation to minimize the number of beds needed? We name the first setting "Human-Oriented" perspective and the second setting "Resource-Oriented" perspective. The nature of these two perspectives is different in the way that the final goal of allocation in the human-oriented perspective is focusing on the individuals being served while in the resource-oriented perspective it is focused on the resources being used. As examples, we can refer to Ala and Chen (2020) who designed an integer programming model for patient scheduling with the goal of minimizing the waiting time of patients in an emergency center in China which can be categorized in the human-oriented perspective while Pei et. al. (2022) used optimization formulation to minimize the number of beds required while meeting a certain threshold for the average waiting time during Covid-19 in a hospital in Wuhan which can be perceived as a resource-oriented perspective.

While most papers picked one perspective, some papers paid the same attention to both perspectives. For example, Alipour-Vaezi, Aghsami, and Jolai (2022) used integer programming techniques to minimize the sum of waiting time of vulnerable patients and service cost in patient scheduling during Covid-19 in in a hospital in Iran, while Mashkani et. Al. (2022) used integer programming techniques to improve a metric defined based on a balance between patients' welfare and hospital administrators' expectations for patient scheduling in operating theatres in Australia.

According to our analysis, 76% of the papers (57 papers) looked at the problem from human-oriented perspective, 11% (10 papers) from resource-oriented perspective and 10% from both perspectives. Among different scopes, the resource-oriented perspective was more popular within disaster and healthcare while all papers in organ transplantation, homelessness and welfare considered the human-oriented perspective (Figure 2).

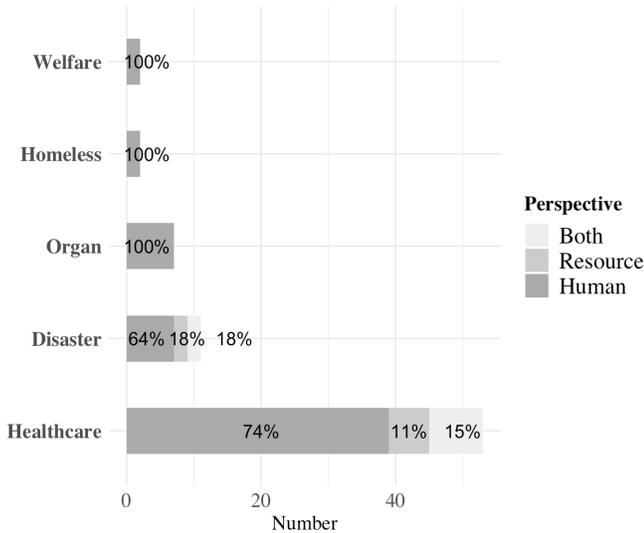

Figure 2: Perspective by Scope

## Approach and Technique: Aggregate vs. Individual or Optimization vs. Prioritization

Apart from the focus of the system which can be on the humans or resources, the overall design of a resource-allocation system can be approached from an individual level or an aggregate level. Considering the problem of organ transplantation from an aggregate perspective, the problem can be formulated as an optimization model with the aim of maximizing the average survival rate of individuals after allocation while from individual perspective it can be prioritizing those who have the most survival chance after allocation.

In the aggregate approach, the problem is usually formulated as an optimization model with an objective while with the individual approach the problem is usually solved with a machine learning algorithm. As examples, in the individual-level we can refer to Börner et. al. (2022) who developed a donor- recipient allocation system based on interpretable neural network techniques to prioritize those who have the highest survival rate after receiving liver while Rempel, Shiell, and Tessier (2021) designed an optimization model with the aim of maximizing the number of survivors during a major maritime disaster which is perceived as an aggregate-level approach. While with the individual approach, the aim of the system is always on the humans rather than resources, with the aggregate approach, the overall aim can be on humans or resources.

According to our analysis at least 60% of the papers used the aggregate approach and optimization techniques for solving a resource allocation problem and among those, integer or linear programming formulation was more popular consisting of 24% of the papers. Among other papers who took individual-approach, decision tree and random forest were more popular consisting of 19% of the papers. A complete breakdown is available in Figure 4. It is worth noting that the methods shown in Figure 4 are not exclusive as some papers might have used different techniques in the process of the design. As an example, we can refer to Rahmattalebi et. al. (2022) who used random forest to predict the circumstances of homeless people after allocating different types of housing and then used mixed-integer optimization techniques to improve the overall rate of exit from homelessness compared to a first come first serve policy.

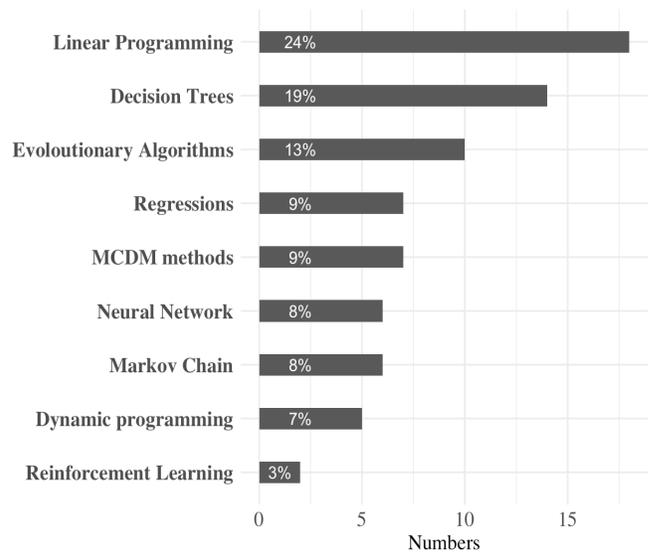

Figure 3: Most Popular Algorithms in the papers

## Optimization goals or Prioritization metrics

Whether the problem is approached from an individual level or an aggregate level, the overall metric on which the resource allocation is based on can be named as the prioritization metric or the optimization goal respectively (Das 2022). While human-oriented versus resource-oriented perspective represents a general direction the system designers take, the final goal of the optimization or the metric chosen for prioritization can be more specific. Table 2 shows different optimization goals or prioritization metrics identified in the papers and categorized into human-oriented versus resource-oriented perspective as described earlier. It is worth noting that we wrote both individual and aggregate metric where there was a correspondence between two. For example, prioritizing those who have the best outcome after allocation in the context of prioritization or maximizing the overall outcome after allocation in the context of optimization (however, we are aware that, as an example and as mentioned by Das (2022) maximizing social welfare might not be the same as choosing those who have the best outcome after allocation).

Previous works in the review of algorithmic allocation of scarce resources have largely focused on the human-oriented perspective looking from the individual level. Das (2022) mentioned three different metrics for prioritization listed as prioritizing those who have the minimum welfare pre-allocation, maximum welfare post-allocation or the greatest increase in welfare.

While we categorized these metrics into human-oriented group, we identified two other metrics that could be included in this group. First, minimizing average waiting time of the individuals which is not the same as the first-come first-serve policy since the amount of time for serving individuals might differ according to the conditions. Second, preference aggregation for which we can refer to Freedman et. al. (2020) who determined patient prioritization ratios using a learning algorithm based on a survey from human respondents for allocation dilemmas for kidney exchange.

In the resource-oriented perspective, we can refer to minimizing cost, maximizing efficiency or utilization, or minimizing operation time. Resource utilization, utility or efficiency as defined by Singhrova (2022) is the fraction of resources allocated or occupied relative to the number of available resources. While individuals would also benefit from optimization as they would wait less in the queue, we categorized this metrics as resource-oriented since the primary focus is on the resources being allocated. We categorized papers based on the metrics introduced in Table 2 in different scopes as shown in Figure 4.

| Human-Oriented | Resource-Oriented |
|---|---|
| Maximizing the overall outcome or prioritizing those who have the best outcome after allocation (Aggregate or individual) | Minimizing cost or the number of resources needed (Aggregate) |
| Maximizing vulnerability index or prioritizing the most vulnerable individuals (Aggregate or Individual) | Maximizing doctors' satisfaction (Aggregate) |
| Maximizing the benefit from allocation or prioritizing those who benefit most (Aggregate or Individual) | Maximising efficiency or utilization (Aggregate) |
| Minimizing the average waiting time (Aggregate) | Minimizing operation time (Aggregate) |
| Preference aggregation (Aggregate or individual) | |

Table 2: Optimization goals or Prioritization metrics identified in the literature.

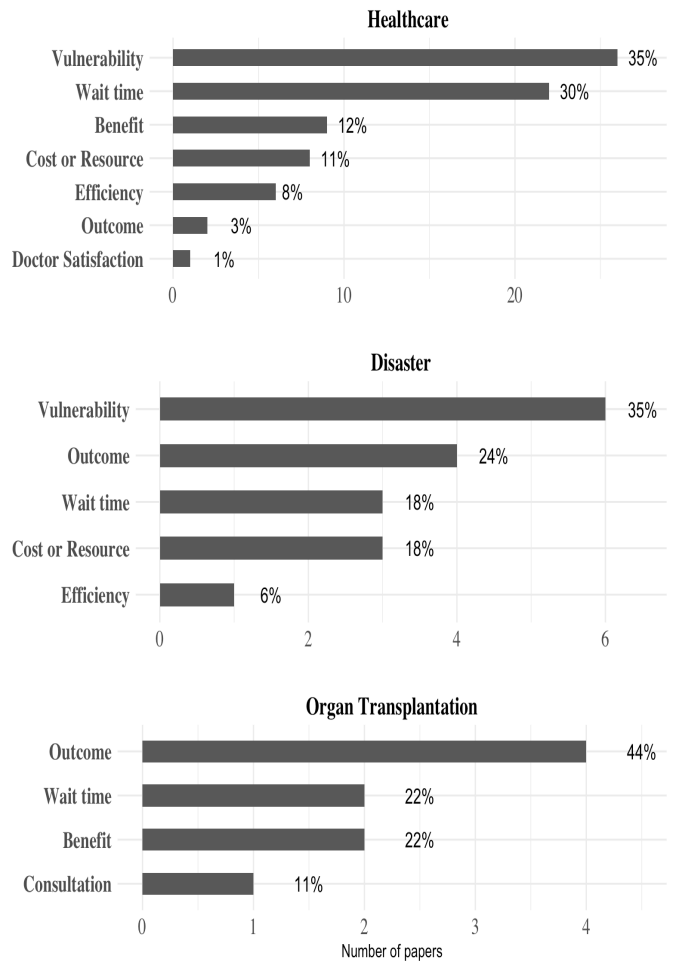

Figure 4: Most popular Metrics within Scopes

Vulnerability was the most popular metric within the healthcare and disaster response areas while outcome was the most observed metric for organ transplantation with at least half of the papers focusing on these metrics in each category. Using terms introduced by Kube et. al. (2022), we can say that healthcare and disaster were more "vulnerability-oriented" compared to organ transplantation which was "outcome-oriented".

Minimizing the waiting time and cost is perhaps more noticed in healthcare compared to other scopes. There were just two papers in the homelessness category among the papers we analyzed that had considered waiting time combined with the benefit or outcome metrics (Dong et. Al. 2021; Rahmattalebi et. Al. 2022) and two papers in the welfare category: Aiken et. al. (2022) used a survey data to train a machine learning algorithms to recognize the patterns of poverty in mobile data and prioritize the poorest individuals for a cash-transfer program during Covid-19 in Togo and Xu, and Liu (2022) used optimization techniques with the goal of maximizing the social welfare for a pension scheme in China.

### Interpretability

The ability for lay humans to understand the decisions algorithms make about them is critically important for notions of justice: for example, in Article 22 of the Europeans Union's General Data Protection Regulation (GDPR), it is mentioned that algorithms should meet the "right to explanation". This might seem challenging especially because interpretability may come at the cost of decreasing efficiency or accuracy if the method of interpretability used involves specifying a simple model. However, Goodman and Flaxman (2017) state that while the "right to explain" may impose large restrictions on the private sector, it can bring good opportunities for designing algorithms which tackle discrimination in the public sector.

In the context of resource allocation, governments and policy makers should be transparent and have a good explanation for why they have prioritized an individual over another. Hence, in the context of our study, designing interpretable systems is important.

We captured whether interpretability was a matter of importance for the authors when designing the system. This concern can include using more interpretable algorithms like decision trees instead of more black box ones like neural networks with the purpose of designing a transparent system or making additional efforts to make a less interpretable algorithm into something more transparent. These efforts usually consist of recognizing the properties which are more critical in the process of selecting individuals. This can happen by choosing the features manually before giving them to the learning algorithm or using an automated algorithm to highlight the most important features in choosing the eligible individuals after learning is done.

According to our analysis just around 20% of the papers showed concern around designing interpretable systems and among them choosing properties before giving them to the learning system and using interpretable algorithms were the most popular techniques (see Figure 5).

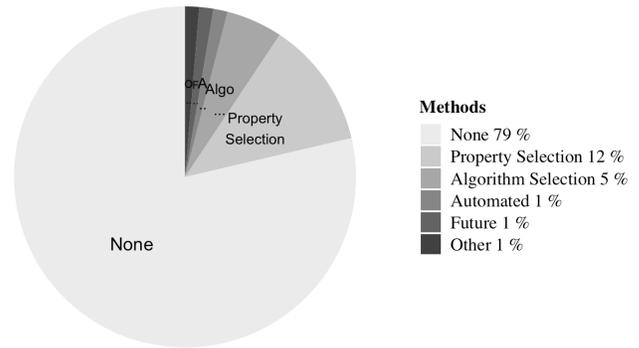

Figure 5: Interpretability in the papers

### Flexibility

As notions of justice in resource allocation might change depending on the scope, time, or territory, the algorithms should have the potential of adapting to changes. In this manner, the output of algorithms should be checked using multiple metrics if it wants to be implemented and useful for a longer period of time or a more general scope.

The design of a flexible tool is important from this perspective in that some criteria might conflict each other, and which criterion should be weighted more in decision making might be case-sensitive and depend on the society's perception of justice. For example, increasing fairness may come at the cost of decreasing efficiency. In this manner, we can refer to Papalexopoulos et. al. (2022) who developed a flexible tool to balance between fairness and efficiency by taking policy makers' preferences for lung transplantation as a case study.

Among the papers we analyzed, 22% of the papers had mentioned some kind of flexibility in their design.

### Ethical Considerations

As mentioned by Das (2022) in the context of allocating scarce resources, two questions arise: "who should be prioritized and why?" The question of who has been considered through the optimization goals and the question of why will be answered through the ethical reason the designers bring for choosing a particular optimization goal and whether they have considered any guidelines, or the society's' preferences, in designing their system. This discussion is important from the perspective that the designers should be aware of the metric they choose for optimization or prioritization and that should be aligned with the perception of the public from the notion of local justice. However, it remains as question whether any universal scope-dependent or independent guidelines can be generated. Vinay, Baumann, and Biller-Adorno (2021) who investigated ethical debate in triage protocols and found points of agreements and disagreements for prioritization, stated that "universality vs. context dependence of principles" is a research area that needs more development.

According to our analysis, only 32% of the papers had considered ethics around the optimization goals they chose by providing a reasoning, taking preferences from experts, or referring to some scope-specific guidelines.

### Bias

In Article 22 of the Europeans Union's General Data Protection Regulation (GDPR), "non-discrimination" is a right of individuals that algorithms should meet. As stated by Altman (2015), "In general, discrimination might be defined as the unfair treatment of an individual because of his or her membership in a particular group, race, or gender".

Although a wide variety of work has criticized algorithms in the public sector for showing discrimination or bias in their operation, among the papers we reviewed, just 4% or three of them had checked their system for bias. It is notable that the widespread public discussions around bias have yet to feed their way into the majority of scientific literature in the area.

### Performance

We captured the performance of algorithms from two different perspectives. First, how well the algorithms can identify or predict the circumstance of individuals before and after allocation (which we refer to as evaluation). Second, comparison of the algorithms to status quo or the current system of allocation which will be described next.

### Evaluation

If the algorithms want to enhance the allocation process, first they should work well in terms of predicting outcome after allocation or identifying the most vulnerable individuals before allocation depending on the prioritization metric. This is usually achieved through evaluating algorithms on a retrospective dataset. Not all the approaches need evaluation. For example, an optimization formulation with the goal of minimizing the waiting time or cost doesn't include any prediction or identification and hence doesn't need any evaluation to validate it based on the retrospective dataset.

However, the performance of the machine learning algorithms on retrospective dataset tells us how reliable the algorithms are.

According to our analysis, the most common evaluation metrics used in the papers were Accuracy, F1, Precision, Recall and AUC respectively. On average, papers could achieve between 82% and 89% on each of these metrics (one may refer to figure A2 in appendix for a box plot of evaluation values).

### Comparison to the Status Quo

If the systems is to be implemented in practice, it should be compared to the status quo or the traditional system of allocation. When the problem is formulated as an optimization model, the metric chosen for comparison is usually the same as the optimization goal of the system.

According to our analysis, survival rate or settling down which is an index for success after allocation, waiting time and utilization were among the most popular metrics reported by the papers. While some papers had just mentioned improvement to the status quo, others reported previous and new values for the metrics. Among the papers that had reported the values, we obtained the percentage change in each of the metrics. By using algorithms, on average waiting time is improved by 49%, utility by 46% and survival rate by 20% compared to the status quo of allocation (one may refer to figure A3 in appendix for a box plot of the values). This is interesting and shows the potential for considerable performance gains across the public sector if such algorithms are used more widely.

## Conclusion and Discussion

In this paper, we analyzed 75 papers in algorithmic resource allocation within healthcare, disaster relief, organ transplantation, homelessness, and welfare. We identified that 76% of them approached the problem from a "Human-Oriented" perspective and the rest from a "Resource-Oriented" perspective or a combination of human and resource. At least 60% of the literature looked at the problem of resource allocation from an aggregate level formulating it as an optimization model and the rest looking at the problem from an individual level using machine learning techniques. Within the healthcare and disaster, "vulnerability-oriented" prioritization metrics and for organ transplantation "outcome-oriented" metrics were more noticed.

Resource-oriented perspective was more popular within disaster and healthcare. Considering the higher cost of disaster and healthcare management and bigger groups of stakeholders and administrations like hospital owners, doctors and nurses, the popularity of resource-oriented perspective in these scopes seems reasonable. However, it remains as a question whether it is ethical to look at the problem of allocating resources to the patients or vulnerable groups from the "Resource-Oriented" perspective.

Designing interpretable systems and considering ethical issues for choosing the prioritization metrics were moderately noticed in the literature with around 20% and 30% of the papers respectively.

It is perhaps promising that algorithms could enhance survival rate and average waiting time of the individuals and utilization of resources compared to the decisions made by humans however, a deficiency of attention to the issues of bias is prevalent in the literature. In addition, whether the resource-oriented perspective is an ethical way of considering algorithmic allocation remains an open question. Whether the systems should be flexible and tested based on different metrics and whether there should be any consensus around the perspective and the metrics chosen needs more investigation.

It is worth noting that although we restricted our time frame to 2020-2023, we tried to capture all the scopes where allocation decision about humans is made. This timeframe, although short, seems reasonable since using algorithms for allocation of resources is a topic which has gained attention in recent years. However, this work can be extended by considering a larger time frame using the lenses introduced here.

While numbers we reached here are subject to change, we hope the overall statistics obtained from the lens of each metric be helpful for researchers and policy makers wanting to address a resource allocation problem.

# Appendix

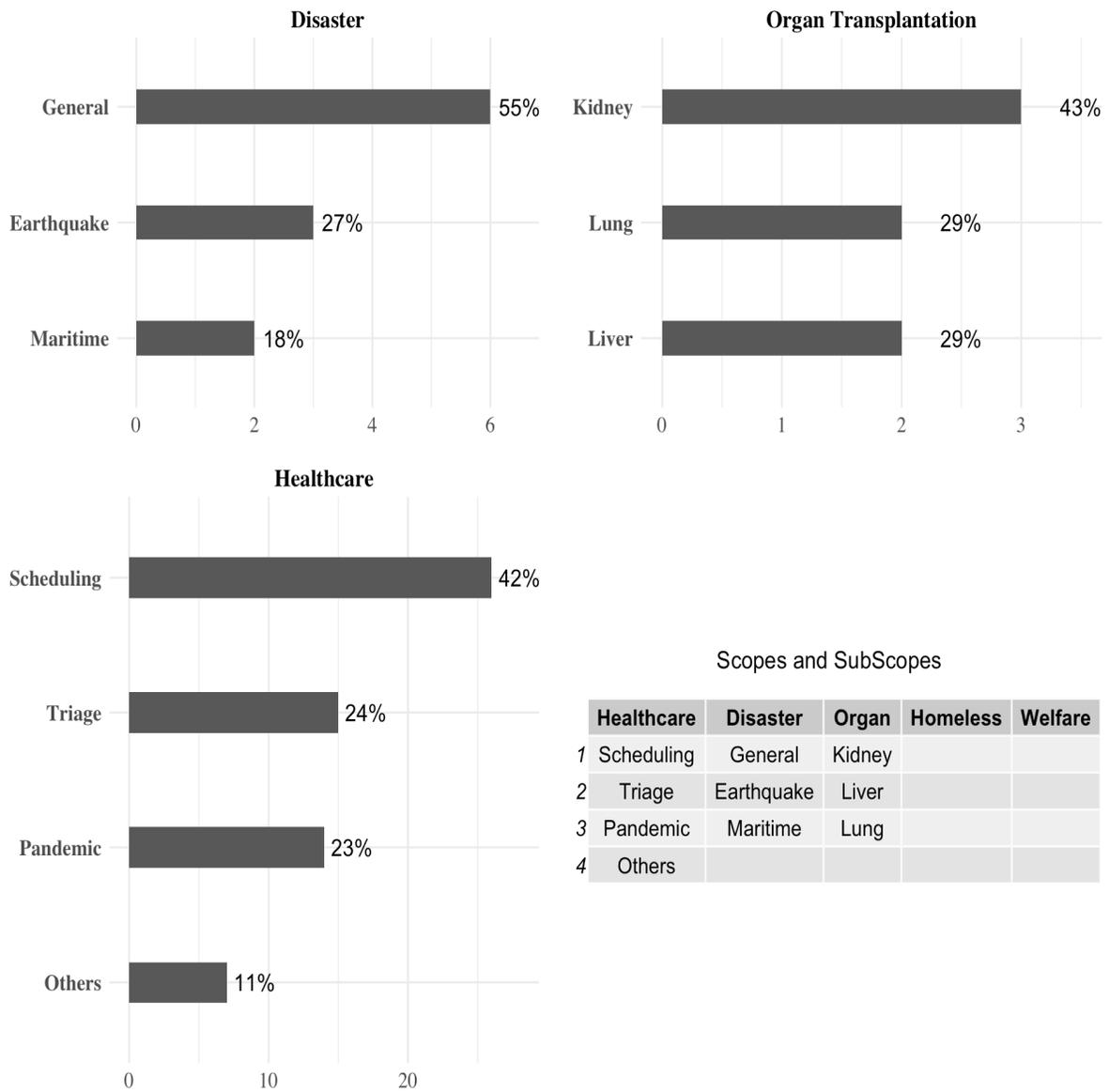

Figure A1: Table and column chart of subcategories for healthcare, organ Transplantation and disaster

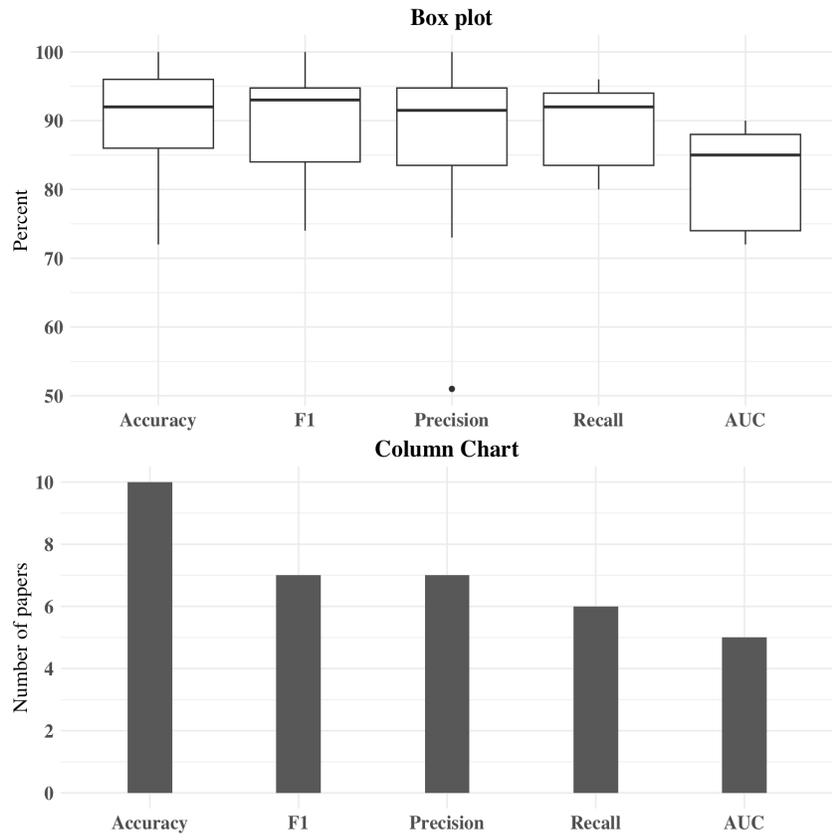

**Figure A2:** Evaluation metrics and box plot of values

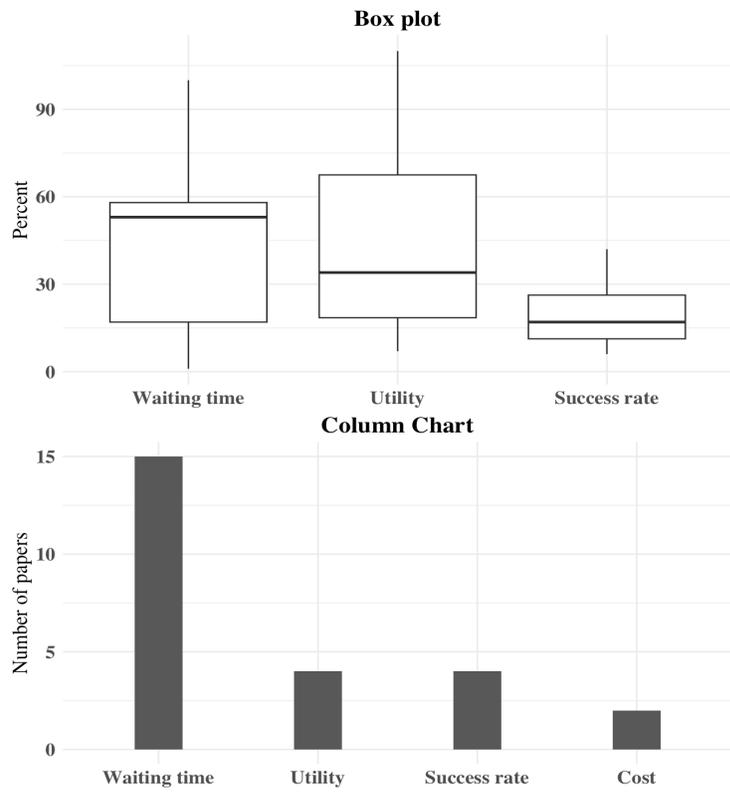

**Figure A3:** Values for comparison and box plot of improvements